# Exotic magnetism in perovskite KOsO$_3$

Jie Chen[1], Hongze Li[1], Javier Gainza[2], Angel Munoz[3], Jose A. Alonso[2], Jue Liu[4], Yu-Sheng Chen[5], Alexei A. Belik,[6] Kazunari Yamaura,[6,7] Jiaming He[1], Xinyu Li[1], John B. Goodenough[1], and J.-S. Zhou[1*]

[1] Materials Science and Engineering program, Mechanical Engineering, University of Texas at Austin, Austin, TX 78712 USA

[2] Instituto de Ciencia de Materiales de Madrid, CSIC, Cantoblanco, E-28049 Madrid, Spain

[3] Universidad Carlos III, Avenida Universidad 30, E-28911, Leganés-Madrid, Spain

[4] Neutron Scattering Division, Oak Ridge National Laboratory, Oak Ridge, Tennessee 37830, USA

[5] NSF's ChemMatCARS, The University of Chicago, Illinois 60437, USA

[6] Research Center for Materials Nanoarchitectonics (MANA), National Institute for Materials Science, Namiki 1-1, Tsukuba, Ibaraki 305-0044, Japan

[7] Graduate School of Chemical Sciences and Engineering, Hokkaido University, North 10 West 8, Kita-ku, Sapporo, Hokkaido 060-0810, Japan

**Abstract**:

A new perovskite KOsO$_3$ has been stabilized under high-pressure and high temperature conditions. It is cubic at 500 K (*Pm*-3*m*) and undergoes subsequent phase transitions to tetragonal at 320 K (*P4/mmm*) and rhombohedral (*R*-3*m*) at 230 K as shown from refining synchrotron X-ray powder diffraction (SXRD) data. The larger orbital overlap integral and the extended wavefunction of 5d electrons in the perovskite KOsO$_3$ allow to explore physics from the regime where Mott and Hund's rule couplings dominate to the state where the multiple interactions are on equal footing. We demonstrate an exotic magnetic ordering phase found by neutron powder diffraction along with physical properties via a suite of measurements including magnetic and transport properties, differential scanning calorimetry, and specific heat, which provide comprehensive information for a system at the crossover from localized to itinerant electronic behavior.

The Bloch wavefunction consists of atomic orbitals and plane waves for describing electrons in a crystal. The Coulomb interaction U between electrons increases the probability for electrons to stay in atomic orbitals, *i.e.* more localized in real space. A sufficiently large U opens a gap near the Fermi energy, which leads to a Mott insulator.[1] Transition-metal perovskite oxides provide a good playground for studying the Mott physics. In addition to the Coulomb interaction, the Hund's coupling plays an important role in determining physical properties in the strongly correlated systems, especially those with electrons in the $t_{2g}$ orbitals in the cubic crystal field.[2] The threshold $U_c$ for the Mott transition becomes the lowest for systems with three electrons in the $t_{2g}$ orbital. In other words, these systems are most likely to be Mott insulators. The cubic perovskite $SrMnO_3$ with the electron configuration $t_{2g}^3 e_g^0$ is a G-type antimagnetic insulator with a Néel temperature $T_N \approx 260$ K.[3] Another family of 3d perovskites $RCrO_3$ having $t_{2g}^3 e_g^0$ is also G-type AFM insulator with $T_N$ in a range from 120 K to 298 K.[4] As orbitals become more extended for 4d electrons, a weaker U in $SrTcO_3$ with $t_{2g}^3 e_g^0$ has been invoked in an argument that the oxide approaches the crossover from localized to itinerant electronic behavior so as to exhibit a high $T_N \sim 1000$ K although it is also a G-type AFM.[5,6] The even more extended 5d electrons with a much weaker U makes $AOs^{5+}O_3$ with $t_{2g}^3 e_g^0$ a good case to test whether the G-type AFM remains stable at the crossover. The orthorhombic $NaOsO_3$ exhibits the G-type AFM with $T_N = 410$ K [7,8]. However, a reduced $T_N$ in $NaOsO_3$ may be attributed to a highly distorted structure which reduces the bandwidth like that in the orthorhombic $RCrO_3$.

The cubic perovskite $KOsO_3$ can be stabilized by quenching the high-pressure phase to ambient condition. The detailed information about the sample preparation can be found in Supplemental Material (SM).[9] The availability of $KOsO_3$ completes the family of $A^{1+}Os^{5+}O_3$ in a broader range of geometric tolerance factor $t = (A-O)/(B-O)\sqrt{2}$. $LiOsO_3$ synthesized under high pressure has a $t = 0.86$ which is too small for the perovskite structure; it crystalizes in the $LiNbO_3$ (LNO) structure.[10] $NaOsO_3$ ($t=0.98$) synthesized under high pressure crystallizes in the orthorhombic perovskite structure. $NaOsO_3$ undergoes a metal-insulator transition at $T_N = 410$ K [7] and the G-type AFM has been determined by neutron diffraction. The transition has been argued as a good example of Slater transition since there is no obvious change in the cell volume on crossing the transition.[8] However, a kink in both *a* and *b* axes of the *Pbnm* orthorhombic cell found in neutron diffraction on crossing the metal-insulator transition has been well mimicked by a DFT calculation along increasing of U, which is associated with a change of the local structural distortion. This work indicates that the Mott physics plays an important role behind the metal-insulator transition.[11] The Néel temperature in a localized electron system increases with increasing the orbital overlap integral (OOI). The bandwidth can be tuned by altering the structural distortion through chemical substitution. In perovskite $AMO_3$, a $t < 1$ is accommodated by the cooperative octahedral-site rotations that lower the structural symmetry and bend the bond angle M-O-M from 180 ° found in the cubic phase for $t=1$. The structural change from the orthorhombic to cubic phase enhances OOI, therefore the bandwidth. The structural change from the orthorhombic $NaOsO_3$ to the cubic $KOsO_3$ broadens up the electron bandwidth. Therefore, studying the evolution of magnetism from $NaOsO_3$ to $KOsO_3$ will illuminate the enigma of magnetism at the crossover.

The structural study by SXRD indicates that KOsO$_3$ undergoes subsequently phase transitions from the cubic C (*Pm-3m*) to the tetragonal at 320 K T (*P4/mmm*) and to the rhombohedral phase R (*R-3m*) at 230 K as shown by lattice parameters versus temperature in Fig.1(a). There is no thermal hysteresis loop in the structure on crossing these phase transitions. The detailed information of the structural refinement can be found in Fig.S1-S4 and Table S1-S6 in SM.[9] There are extremely small local structural distortions associated with the two phase transitions. The Os-O bond length along the *c* axis is slightly smaller than that in the *a-b* plane in the T phase and the O-Os-O bond angle deviates slightly from 90 ° in the R phase. Neutron powder diffraction (ND) has been performed in the same temperature range. Although the very tiny peak splitting associated with the phase transitions detected by SXRD cannot be distinguished in the ND, the comparison of patterns between SXRD and ND at 500 K in Fig.1(b) unambiguously illustrate the magnetic diffractions, which indicates a magnetically ordered phase in KOsO$_3$ at least at 500 K. Further justifications that the extra peaks from ND are from magnetic ordering are given in Fig.S5-S9 of SM.[9]

We start with the magnetic structure in the cubic phase at 500 K. The initial fitting with magnetic structures having **k**=0 fails. The possible magnetic structures are given by the basis vectors of the irreducible representations of the little group G$_k$, which is formed by the symmetry elements of the space group *Pm-3m* that leave **k** invariant. For the cubic KOsO$_3$, there is only one Os atom per unit cell, so the basis vectors can only establish some relationships between the m$_x$, m$_y$ and m$_z$ components of the magnetic moment. For **k$_1$** = (k$_x$, k$_y$, k$_z$), the little group G$_{k1}$ is only formed by the identity, in which there is only one irreducible representation and there is no relationship between the m$_x$, m$_y$ and m$_z$ components.[12] The basis vectors of three phases are presented in Table S7.[9] A good agreement between the observed and calculated neutron diffraction patterns in bank1 at 500 K based on a model of spin structure is obtained and shown in Fig.1(b), Fig.1(c) for a zoom-in plot at magnetic peaks, and Fig.S10 for the fitting results in bank2. The magnetic structure with a sinusoidally modulated moment and the maximum moment M~1.5 μ$_B$ is incommensurate to the lattice; a schematic plot of the spin structure projected in the *a-c* plane of the cubic cell is given in an inset of Fig.1 and the magnetic moment direction is shown in Fig.S11.[9] The important parameters of fitting the neutron diffraction patterns at 500 K are shown in Table S8.[9] The model of magnetic structure in T and R phase is similar to that in C phase, but has slightly different **k**. The fitting results with the model are shown in Fig.S10 and fitting parameters are in Table S9.[9]

The non-linear M-H curves in Fig.2(a) from 300 K to 680 K and the hysteresis loop in Fig.S12 of SM [9] confirm the magnetically ordered phase. The magnetization at H = 2T and room temperature is comparable to that in NaOsO$_3$ [7] in which the small spontaneous magnetization was attributed to the spin canting allowed by the G-type AFM in the *Pbnm* perovskite structure. The sample starts to decompose at T > 600 K from the magnetization measurement, which prevents us from determining T$_N$ accurately. However, extrapolating the curve of coercive force versus temperature (Fig.S12 of SM [9]) to higher temperatures can give a rough estimation of T$_N$ ~ 800 ± 100 K. The structural phase transitions at 230 K and 320 K correspond to anomalies at these temperatures in the temperature dependence of magnetization in Fig.2(b) which also has the same magnitude as that of NaOsO$_3$ below T$_N$.[7] Since all the crystal structural of three phases found in $KOsO_3$ are not compatible with spin canting, the nonlinear magnetization must be attributed to a weak ferromagnetism, which will be further elaborated below.

Measurements of resistivity in Fig.2(c) confirm that $KOsO_3$ is a poor metal with an overall resistivity comparable with that of the metallic phase at $T > T_{IM}$ in $NaOsO_3$. However, the subsequent phase transitions from C to T to R phase cause jumps in the resistivity. There are corresponding anomalies in the temperature dependence of thermoelectric power S in Fig.2 (e); S drops abruptly and changes sign on cooling from C phase to T phase whereas the magnitude of |S| reduces through the transition from T phase to R phase. The overall magnitude and the temperature dependence of S are consistent with a metal with the Fermi level located in the middle of π* symmetry band shown in Fig.S14.[9,13] The specific heat result in Fig.2(d) can fit well to the formula of Debye-Einstein model at all temperatures and $C_p(T) = \gamma T + \beta T^3$ at low temperatures. A $\gamma = 2.54$ mJ/mol $K^2$ from the fitting result is the smallest one for a metal; a $\gamma = 4.25$ mJ/mol $K^2$ is obtained from the band structure in Fig.S13 [9]. The λ-shape anomaly in $C_p(T)$, typical for the second-order phase transition can be clearly seen at 230 K and 320 K. The entropy change $\Delta S=1.7$ J mole$^{-1}$ K$^{-1}$ associated with the C-T phase transition and the results of DSC measurement from 100-500 K are given in Fig.S15, S16 of SM.[9] Comparing with a $\Delta S=0.6$ J mole$^{-1}$K$^{-1}$ associated with a pure structural transition in the perovskite $LaGaO_3$,[14] a huge $\Delta S$ at the C-T transition must be due to some change of electronic degree of freedom.

The spin canting due to the Dzyaloshinskii-Moriya (DM) theory occurs only in some structural systems with lower symmetries, such as the orthorhombic *Pbnm* perovskite.[15][16] The abrupt cancellation of canted moment from the orthorhombic to cubic perovskite in G-type AFM $Ca_{1-x}Sr_xMnO_3$ is a good example.[3] None of the three phases found in $KOsO_3$ is compatible with canted spins. The coexistence of the magnetic structure with a sinusoidally modulated moment and conducting electrons in the perovskite $KOsO_3$ fits the prediction of the Hubbard Hamiltonian at the crossover.[17] In this theory, the effect of U is not to split a band into UHB and LHB in one step, but to transfer the spectral weight around the Fermi energy $E_F$ of free electrons into the Hubbard bands continuously. At the crossover, there are coherent electrons near $E_F$ and electrons in LHB responsible for static magnetic moments. The nonlinear magnetization in the $KOsO_3$ can be well attributed to the weak itinerant electron ferromagnetism [18][15] contributed from the coherent electrons at $E_F$.

The evolution of magnetism in transition-metal perovskites can be demonstrated along pathways of either the structural distortions or the broadening of the d orbital distribution function from 3d to 4d to 5d orbitals for systems with the $t_{2g}^3 e_g^0$ configuration. In the perovskite structure, *b* of OOI is related to the orthorhombic structural distortion, specifically, the M-O-M bond angle $\psi=180-\phi$ by the relationship of the bandwidth $W \sim b \sim \cos\phi$. The transition from the orthorhombic phase ($\psi=180-\phi$) to the cubic phase ($\psi=180$) enhances *W* (given a constant M-O bond length). The magnetic transition temperature, therefore, can be linked to the structural parameter through the perturbation expression of superexchange interaction (SEI),[19] $T_N \sim b^2/U \sim \cos^2\phi/U$, which implies lines with a nearly identical slope for different perovskite systems with localized electrons in the plot of $\ln(T_N)$ versus $\ln(\cos\phi)$.

The systems having $t_{2g}^3e_g^0$ give the highest possible spin state available for comparing the change of magnetic properties in perovskites with 3d, 4d, and 5d transition metals. In the 3d oxides, there are two families of $RCrO_3$[4] and $Ca_{1-x}Sr_xMnO_3$ suitable.[3] The G-type antiferromagnetic ordering identified in these two families is in line with the rules for SEI. The rare-earth substitution in the orthorhombic perovskite $RCrO_3$ changes the Cr-O-Cr bond angle from 145º to 160 º for R= Lu to La.[4] Data points from all members in this perovskite family fall almost perfectly on a line in the plot of $ln(T_N)$ versus $ln(cos\phi)$ in Fig.3(a). For the series of $Ca_{1-x}Sr_xMnO_3$, a line with nearly the same slope as that for $RCrO_3$ can be obtained if it is drawn between the spots for $CaMnO_3$ and $SrMnO_3$. A slightly lower $T_N$ relative to the line for the $Ca_{1-x}Sr_xMnO_3$ with 0 < x <1 can be well accounted for by the size variance effect.[20][21] This plot confirms that the perturbation expression [19]of SEI is applicable to describe the magnetic interactions in these 3d perovskite systems with localized electrons. Moreover, the structural change from the orthorhombic phase to the cubic phase adds a clear restriction on the spin structure. The canted spin structure is allowed in the orthorhombic structure based on either the DMI [22] or the single ion anisotropy (SIA),[23] which is represented by a spontaneous magnetization along the canted spin direction and a diverge increase of M(T) on cooling through $T_N$ in $CaMnO_3$. These features disappear in the cubic $SrMnO_3$.[3][24]

The 4d perovskites $CaTcO_3$ [25] and $SrTcO_3$ [5] show the highest magnetic transition temperature in transition metal oxides due to the much extended 4d orbitals. The colinear G-type spin structure found in $CaTcO_3$ and $SrTcO_3$ and the positive slope of a line connecting $CaTcO_3$ and $SrTcO_3$ in the plot of $ln(T_N)$ versus $ln(cos\phi)$ in Fig.3(a) are consistent with the scenario of localized electrons in these oxides. However, a reduced magnetic moment relative to the 3d perovskites in Fig 3(b) reflects the extended wavefunction for the 4d electrons.

While an even higher $T_N$ is expected in $AOs^{5+}O_3$ (A = alkaline elements) because of a larger OOI through even more extended 5d orbitals and a weaker U based on SEI, the strong spin-orbit coupling (SOC) associated with Os (z = 78) compared with Tc (z = 43) competes with the spin-spin interaction.[26,27] The SOC effect should be negligible for the L-S coupling as the orbital momentum is quenched for $Os^{5+}$( $t_{2g}^3e_g^0$). However, the j-j coupling which leads to a $J_{eff}$ =1/2 state in $Sr_2IrO_4$,[28] still places the SOC on the same footing with the spin-spin interaction in these 5d oxides so as to lower the magnetic transition temperature. A large change of the bond angle Os-O-Os from the orthorhombic $NaOsO_3$ to the cubic $KOsO_3$ separates these two perovskites in the plot of $ln(T_N)$ versus $ln(cos\phi)$ in Fig.3(a). The nearly same slope for the line connection $NaOsO_3$ and $KOsO_3$ as those for $RCrO_3$ and $AEMnO_3$ in Fig.3(a) may be an accident. But the positive slope for the $T_N$ change between $NaOsO_3$ and $KOsO_3$ in the plot and the G-type magnetic ordering in $NaOsO_3$ fit the picture of localized electrons in $NaOsO_3$. The transport properties and the specific heat in Fig.2(c-f) clearly place $KOsO_3$ on the itinerant electron side of the crossover. Two important findings from this work: (1) On top of the moment modulation, the spin arrangement in the *ac* plane creates the interfaces of AF coupling that is incommensurate with the lattice. The incommensurate spin structures reported in the literature refer to the spiral spins that travel in the lattice transversely.[29][30][31][32][33][34] The unprecedented incommensurate moment variation found in $KOsO_3$ is an example of electronic structure at the crossover. The average moment of the sinusoid wave of the moment at Os sites in $KOsO_3$ is nearly the same as that in homogeneous $NaOsO_3$. (2) Whereas there is a sharp change of the magnetic structure from the G-type AFM found in the orthorhombic $NaOsO_3$ to the incommensurate spins in the cubic $KOsO_3$, both perovskites show the colinear spin arrangement. The orthorhombic distortion in a perovskite structure allows or induces [35] the spin canting that leads to a spontaneous magnetization in the G, A, and C types of AFM.[23] The canted spins in $NaOsO_3$ account for the non-linear magnetization. However, the spontaneous magnetization found in the $KOsO_3$ in Fig.2(a) and the clear coercive force plotted in Fig.S12 comes from the weak itinerant electron ferromagnetism in coherent electronic states near $E_F$.

The M-O bonding covalency or the *d-p* hybridization in the transition-metal oxide perovskites reduces the magnetic moment in spin ordered structures. Fig.3(b) displays the change of magnetic moment in four families of perovskites with 3d, 4d, and 5d electrons. The 3d perovskites exhibit the saturation moments M = 2.9-2.5 $\mu_B$ that are close to 3$\mu_B$ for the spin-only value for localized electrons. A reduction to M = 2.1 $\mu_B$ in $SrTcO_3$ has been accounted for by the bonding covalency in a first-principle calculation.[36] An even smaller M = 1.09 $\mu_B$ in $NaOsO_3$ is consistent with the enhancement of the bonding covalency for 5d electrons. The crystal field on $t_2$ orbitals become $\psi_t = N_t(f_t - \lambda_\pi \varphi_\pi)$ after including the covalent mixing with the anion; where $N_t$ is the normalization constant, $f_t$ is d orbital, $\varphi_\pi$ is a $t_2$-symmetrized $p_\pi$ orbital. The moment reduction of the d orbital is caused by the covalent mixing $\lambda_\pi = b^{ca}/(E_t-E_\pi)$, where $b^{ca}$ is the orbital overlap integral in M-O bonds. For the orthorhombic perovskites, the overlap integral over the M-O-M bond becomes $b_\pi = \varepsilon_\pi \lambda_\pi^2 cos\phi$. Given the nearly constant M-O bond length for the perovskites discussed here, the increase of $b_\pi$ comes from a monotonic reduction of $\Delta = E_t - E_\pi$ and a significant increase of $b^{ca}$ due to the change of radial distribution function of *d* orbital from 3d to 4d to 5d. The change of $b_\pi$ between $NaOsO_3$ and $KOsO_3$ is caused by the structural factor $cos\phi$. This change plus a sufficiently large $\lambda_\pi$ as illustrated by a strong Os $t_{2g}$:$O_{2p}$ hybridization as indicated in the DFT calculation of Fig.S14 for $KOsO_3$ makes it the first perovskite with the $t^3e^0$ configuration at the crossover. The schematic plot of $T_N$ versus $b_\pi$ in Fig.3(c) builds in the effects on the bandwidth enhancement from both the structure and the radial distribution function of different orbitals.

In conclusion, the electron bandwidth in the perovskite systems can be varied by the structural distortion and the covalent mixing between the *d-p* orbitals which is sensitive to orbital distribution function. The synthesis of cubic $KOsO_3$ completes a group of perovskites with the highest spin state $t_{2g}^3 e_g^0$ crossing 3d, 4d, and 5d of transition metals, which enables the study of evolution of magnetism. The Hund's coupling effect makes most of these perovskites a Mott insulator. Whereas the antiferromagnetic phase in the 3d perovskites can be described by the Heisenberg Hamiltonian beautifully, much enhanced $T_N$ and reduced magnetic moment in the 4d and 5d perovskites fit predictions of the simulation of the Hubbard Hamiltonian. $KOsO_3$ has the highest electron bandwidth which places it at the crossover from localized to itinerant electronic behavior. The incommensurate magnetic moment revealed by neutron diffraction and anomalous physical properties in connection with the unusual structural changes advance our knowledge of a system at the crossover.

**Acknowledgments**

This research was primarily supported by the National Science Foundation through the Center for Dynamics and Control of Materials: an NSF MRSEC under Cooperative Agreement No. DMR-1720595 and DMR-2308817. JBG was supported by the Welch Foundation (F-1066). JG thanks Ministerio de Ciencia e Innovación (MICINN) for granting the contract PRE2018-083398. JAA thanks the Spanish Ministry of Science and Innovation for granting the project number: PID2021-122477OB-I00.  NSF's ChemMatCARS, Sector 15 at the Advanced Photon Source (APS), Argonne National Laboratory (ANL) is supported by the Divisions of Chemistry (CHE) and Materials Research (DMR), National Science Foundation, under grant number NSF/CHE- 1834750. Use of APS, an Office of Science User Facility operated for the U.S. Department of Energy (DOE) Office of Science by ANL, was supported by the U.S. DOE under Contract No. DE-AC02-06CH11357. Neutron powder diffraction measurements used resources at the Spallation Neutron Source (NOMAD instrument), a DOE Office of Science User Facility operated by the Oak Ridge National Laboratory. MANA is funded by MEXT's WPI, Japan. Synchrotron radiation at SPring-8 (BL02B2) was used under Japan Synchrotron Radiation Research Institute approval (2023A2361, 2023B1676). All simulations were performed on the high-performance computational clusters of Texas Advanced Computing Center (TACC). JSZ is grateful to the enlightening discussion with Allan MacDonald.

*Corresponding author: jszhou@mail.utexas.edu

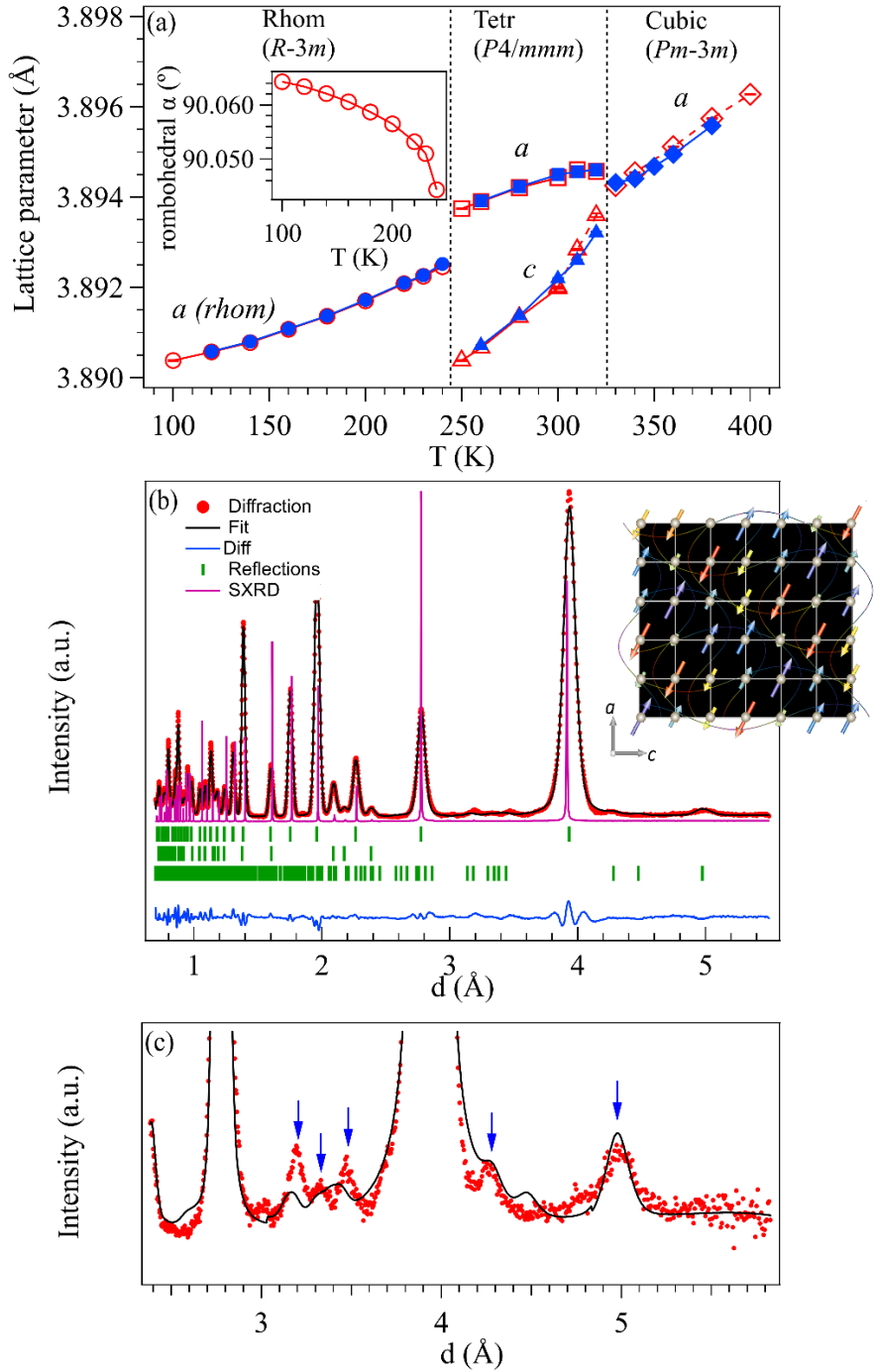

Fig.1 (a) Lattice parameters versus temperature for KOsO$_3$ obtained by Rietveld refinement of the synchrotron XRD profile ($\lambda = 0.61928$ Å) collected at warming and cooling processes. The data acquired from warming up is denoted by red markers, while data from cooling down is represented by blue markers. (b) Neutron powder diffraction (ND), synchrotron X-ray diffraction at 500 K, and the fitting results of ND (the reflections from top to bottom represent nuclear structure of KOsO$_3$, nuclear structure of Os, and magnetic structure of KOsO$_3$); the inset is the spin structure from the fitting. (c) The zoom-in plot of ND and the simulation.

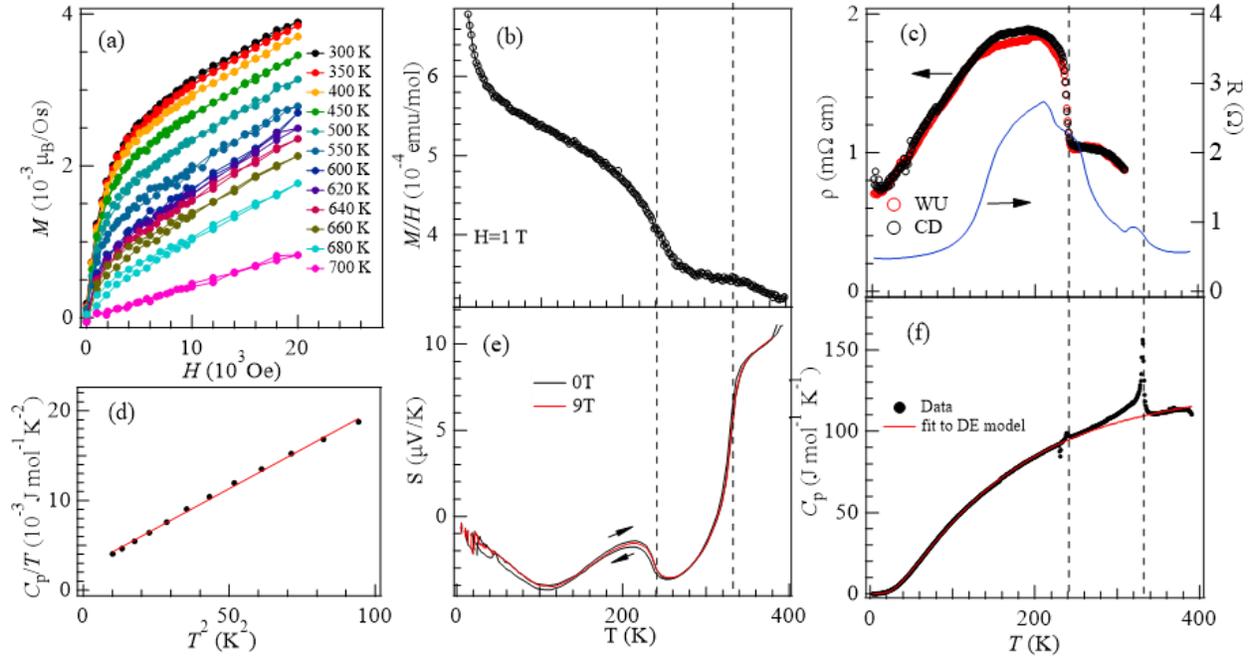

Fig.2 (a) Field dependence of the magnetization of $KOsO_3$ at different temperatures; temperature dependences of (b) the magnetization, (c) the resistivity, (d, f) the specific heat, and (e) the thermoelectric power. The symbols in (c) represent the results measured on a cluster of $KOsO_3$ crystal (WU: warm up; CD: cool down); the solid line shows the result of pressed powder sample in a diamond anvil cell.

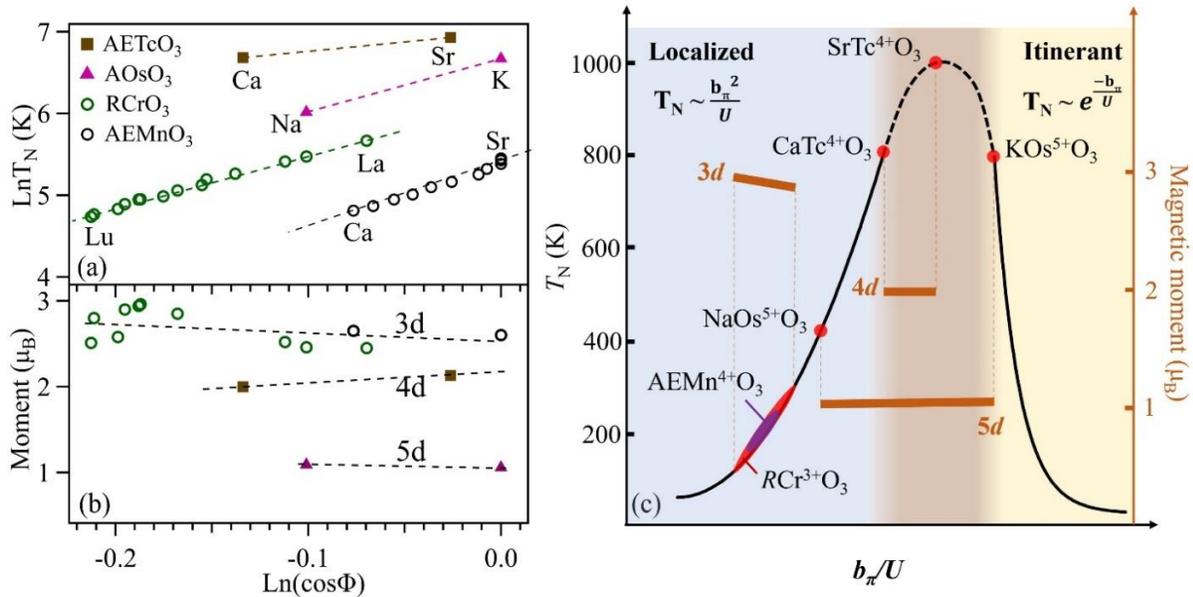

Fig.3 (a,b) The correlation between the Néel temperature $T_N$ and the magnetic moment at 4.2 K and the structural distortion in the perovskite structure for the 3d, 4d, and 5d perovskites with the $t_{2g}^3 e_g^0$ electronic configuration. The angle $\phi$ measures the deviation of bond angle M-O-M from 180º in the orthorhombic perovskite structure. A in the formula stands for alkaline elements, AE for alkaline earth elements, and R for rare earth elements. The magnetic moments in $RCrO_3$ are from ref [37]; the moment of $SrMnO_3$ at 77 K is from ref [24], which is converted to the moment at 4.2 K following the Brillouin function (J=3/2); the moment of $Ca_{1-x}Sr_xMnO_3$ is from ref [3]; the moment of $CaTcO_3$ is from ref [25] ; the moment of $SrTcO_3$ is from ref [5]; the moment of $NaOsO_3$ at 200 K is from ref [8], which is converted to the moment at 4.2 K following the Brillouin function (J=3/2). (c) A schematic phase diagram shows the evolution of the magnetic transition temperature as a function of electron bandwidth from localized electrons to itinerant electrons.